\documentstyle[11pt]{article}
\begin{document}

\begin{titlepage}
\centerline{\LARGE Numerical Renormalization Group
at Criticality }
\vskip 15pt
\centerline{ T. Nishino$^{1}$, K. Okunishi$^{2}$,
and M. Kikuchi$^{2}$ }
\centerline{\sl $1$ Department of Physics,
Graduate School of Science, }
\centerline{\sl Tohoku University, Sendai 980-77, Japan }
\centerline{\sl $2$ Department of Physics,
Graduate  School of Science, }
\centerline{\sl Osaka University,  Toyonaka,
Osaka 560, Japan }
\vskip 20pt

\begin{abstract}
\setlength{\baselineskip}{24pt}
We apply a recently developed numerical renormalization
group, the corner-transfer-matrix renormalization group
(CTMRG), to 2D classical lattice models at their critical
temperatures. It is shown that the combination of CTMRG
and the finite-size scaling analysis gives two independent
critical exponents.
\end{abstract}
\vskip 1.0cm

\begin{itemize}
\item[\bf PACS codes:] 64.60.Fr, 75.10.Hk, 02.70.-c
\item[\bf Key Words:] Renormalization Group, Finite Size
Scaling, Critical Phenomena
\end{itemize}

\begin{itemize}
\item[\bf Address:] Kouichi Okunishi c/o Macoto Kikuchi
\item[] Physics Department, Graduate School of Science,
\item[] Osaka University, Toyonaka, Osaka 560, JAPAN
\item[\bf Phone:] +81-6-850-5349, {\bf Fax:} +81-6-845-0518
\item[\bf e-mail:] okunishi@godzilla.phys.sci.osaka-u.ac.jp
\end{itemize}
\end{titlepage}

The renormalization group is a basic concept in statistical
physics \cite{Kad,Wi}. The real-space renormalization group
(RSRG) has been applied widely to critical phenomena
\cite{Kad,Rs}. A resent progress in RSRG is the development
of the density matrix renormalization group (DMRG) by White
\cite{Wh1,Wh2}. The DMRG was originally formulated for
one-dimensional (1D) quantum models, and was shown to be
applicable to 2D classical systems as well\cite{Ni}.
Recently, \"Ostlund and Rommer found a variational principle
hiding behind DMRG \cite{Os}. They showed that the ground
state wave function obtained by DMRG is in a form of a matrix
product\cite{Zi,Zi2,Zi3}. This variational background enables
us to accelerate the numerical calculation of DMRG
\cite{Os,No2}. Quite recently Mart\'in-Delgado and Sierra
have obtained a unified analytic formulation of conventional
RSRG and DMRG \cite{Si1,Si2}

Baxter established another RSRG for 2D classical systems
by using the corner transfer matrix (CTM)
\cite{Bx1,Bx2,Bx3}. His method is known as a natural
extension of variational methods, such as Kramers-Wannier
approximation \cite{Kw}, Kikuchi's approximation \cite{Ki},
and the cluster variational method \cite{Cv}. It should be
noted that  Baxter's method and White's DMRG method have
many features in common. In particular, both are RSRG based
on a variational principle. On the basis of this fact,
Nishino and Okunishi have formulated a new renormalization
group procedure for 2D classical lattice models
\cite{No,No2}, which we will refer `corner-transfer-matrix
renormalization group (CTMRG)' in the following. It has been
confirmed that the CTMRG precisely determines thermodynamic
functions of 2D classical systems in the thermodynamic limit.
For example, calculated internal energy of the square-lattice
Ising model is -0.70704 at $T_c$ \cite{No}, whose numerical
precision is comparable to a resent Monte Calro result
\cite{So}.

In this letter, we show that the CTMRG method determines not
only the one-point functions, but also critical exponents
of 2D classical spin systems. We apply CTMRG to finite size
systems at their critical temperatures, and calculate
one-point functions at the center of the system. The finite
size corrections to these one-point functions are found to
obey the finite-size scaling (FSS) behaviors determined by
two independent critical exponents \cite{Fi,Ba}.

The CTMRG is a numerical method which can evaluate the
partition function of finite size systems. Consider a square
cluster of a classical spin system whose linear dimension $N$
is an odd integer. The cluster consists of four sub-clusters
of the size $(N+1)/2$, which are named `corners' according to
Baxter; The partition function is calculated as the trace of
a matrix $\rho_{(N+1)/2} = \left( A_{(N+1)/2} \right)^4$,
where $A_{(N+1)/2}$ is so called `corner transfer matrix'
(CTM) which transfers column spins into row spins \cite{Bx3}.
It, however, is difficult to deal with $A_{(N+1)/2}$ exactly
for large $N$ because the dimension of the matrix increases
rapidly with $N$. {\it The point of CTMRG is, that the matrix
$\rho$ is identical to the density matrix that appears in
White's DMRG} \cite{Wh1,Wh2}. This correspondence enables us
to transform $A_{(N+1)/2}$ into a renormalized one ${\bar
A}_{(N+1)/2}$. We determine the dimension $m$ of the reduced
matrix ${\bar A}_{(N+1)/2}$ so that we can deal with ${\bar
A}_{(N+1)/2}$ in realistic numerical calculations. By using a
recursive relation between ${\bar A}_{(N+1)/2}$ and ${\bar
A}_{(N+3)/2}$, we can increase the linear size of the corner
one by one, and obtain ${\bar A}_{(N+1)/2}$ from $N = 3$ to
$\infty$ one after another; this recursive procedure
corresponds to the `infinite chain method' in DMRG
\cite{Wh1,Wh2}. The renormalized CTM ${\bar A}_{(N+1)/2}$
thus obtained gives the approximate partition function as
${\bar Z}_N = {\rm Tr} \, {\bar A}_{(N+1)/2}$. Because of the
variational nature of the CTMRG, ${\bar Z}_N$ gives a
lower-bound for the exact partition function $Z_N = {\rm Tr}
A_{(N+1)/2}$.

The error in the partition function $\delta Z_N \equiv Z_N -
{\bar Z}_N$ is related to two characteristic length scales.
In what follows, we denote the largest and the second largest
eigenvalues of the row-to-row transfer matrix as
$\Lambda_0(N,m)$ and $\Lambda_1(N,m)$ respectively, which can
be calculated by CTMRG. The first characteristic scale is
the correct {\it correlation length} $\xi_N$ for a finite system
defined as follows:
\begin{equation}
1/\xi_N = \log\frac{\Lambda_0(N,\infty)}{\Lambda_1(N,\infty)}.
\end{equation}
According to the finite-size scaling theory, $1/\xi_N$ is a
decreasing function of $N$ as $\sim 1/N$ at the critical
temperature $T = T_c$. The other scale is an {\it effective}
correlation length $\xi(m)$ for finite $m$ in the
thermodynamic limit:
\begin{equation}
1/\xi(m) = \log\frac{\Lambda_0(\infty,m)}{\Lambda_1(\infty,m)},
\end{equation}
This latter scale appears due to the restriction imposed on
the size of the matrix ${\bar A}_{(N+1)/2}$. As long as the
condition $\xi_N \ll \xi(m)$ is satisfied, ${\bar Z}_N$ is a
good approximation for $Z_N$. But in the opposite case $\xi_N
\gg \xi(m)$, ${\bar Z}_N$ will be appreciably smaller than
$Z_N$. Therefore, we expect to observe crossover from the
correct system-size dependence of the partition function to
finite-$m$ behavior. For off-critical case, we can make the
condition $\xi_N \ll \xi(m)$ satisfied by taking large
enough $m$ \cite{No}. For the critical case, on the other
hand, we have to deal with the above crossover properly.

In addition to the partition function, we can also calculate
the local energy $E(N)$ and the order parameter $M(N)$ at the
center of the square clusters of the linear size $N$\cite{No}.
According to the above discussion on the crossover,
we expect that the following two-parameter finite-size
scaling form \cite{Fi,Ba}   holds for these thermodynamic
functions at $T_c$: The $N$ dependence of $E(N)$ is
\begin{equation}
E(N) - E(\infty) = N^{1/\nu -d} f(\xi(m)/N),
\end{equation}
where $\nu$ is one of the critical exponents (the correlation
length exponent) and $d=2$ is the spatial dimensionality.
In addition to the leading finite-size scaling form $\sim
N^{1/\nu -d}$, we introduced an unknown scaling function $f$
which describes the crossover. The asymptotic form of the
scaling function will be $f(x \rightarrow 0) \sim x^{-1/\nu
+d}$ and $f(x \rightarrow\infty) \sim const.$. Similarly, the
size dependence of the local order parameter $M(N)$ is
expected to be
\begin{equation}
M(N) = N^{-(d-2+\eta)/2} g(\xi(m)/N)
\end{equation}
with another set of critical exponent $\eta$ (the anomalous
dimension of the spin) and a scaling function $g$ introduced;
The asymptotic behavior of $g$ will be $g(x \rightarrow 0)
\sim x^{(d-2+\eta)/2}$ and $g(x \rightarrow\infty) \sim
const.$. Assuming these two scaling form, we can
estimate two independent critical exponents $\nu$ and $\eta$
as well as $E(\infty)$ from the data given by the CTMRG.

As examples of the 2D classical lattice models, we deal with
Ising model and 3-state Potts model on the square lattice.
The exact values of the critical exponents appearing in the
scaling forms are $\nu=1$ and $\eta=1/4$ for the Ising model
and $\nu=6/5$ and $\eta=4/15$ for the Potts model. In order
to calculate $M(N)$, we impose the ferromagnetic boundary
conditions, that is, all the spins at the boundary of the
clusters take the same value.

Figure 1 shows the local energy  $E(N) = \bigl\langle
\sigma_i \sigma_{i+1} \bigr\rangle$  of the Ising model
against $1/N$ at $T_c = 2.269185314$. We set $m = 148$; the
result for $m = 4$ is also shown for comparison. The linear
dependence of $E(N)$ on $1/N$ is observed in a wide range of
$1/N$, which is consistent with the leading FSS behavior in
Eq.~(3) with $\nu=1$. In fact, the least-square fitting of
the data  in the range $21 \le N \le 401$ to Eq.~(3) gives
$\nu = 0.993$ and $E(\infty) = 0.70704$, which is close to
the exact value $E(\infty)= 1 / \sqrt{2} =0.70711.$ Figure 2
shows the spin polarization $M(N) = \bigl\langle \sigma
\bigr\rangle$ for $m = 4, 10,$ and $148$.  The $N$ dependence
is well expressed by $N^{-1/8}$ as we expected. The crossover
effect discussed above is clearly seen in the figure: $M(N)$
for different $m$ deviate from $N^{-1/8}$ behavior one by one
when $N$ increases. By the least-square fitting in the range
$21 \le N \le 401$ we get $\eta = 0.2506$.

Now we check the full two-parameter scaling form for $E(N)$
taking the crossover effect expressed by the scaling function
$f$ into account. In figure 3 we plot  $N \left\{E(N)
-E(\infty)\right\}$ against $N/\xi(m)$ for $m = 4 \sim 13$;
we choose relatively small $m$ in order to observe small
$\xi(m)/N$ region. All the data are really collapse into a
single curve, and thus the two-parameter scaling assumption,
Eq.~(3), is justified.

Figures 4 and 5 show the local energy $E(N) = \bigl\langle
\delta( \sigma_i \sigma_{i+1} ) \bigr\rangle$ and the local
order parameter $M(N) = \bigl\langle \delta(0,\sigma)
\bigr\rangle$, respectively, of the 3-state Potts model at
$T_c = 1.989944809$. Again the expected $N$ dependence,
$E(N)-E(\infty ) \sim N^{-4/5}$ and $M(N) \sim N^{-2/15}$,
are observed. The calculated critical exponent $\nu$ is 0.830
in the range $5 \le N \le 31$, and is 0.809 in the range $4
\le N \le 201$. For the exponent $\eta$ we get $0.266$ in the
range $5 \le N \le 31$, and $0.267$ in the range $5 \le N \le
201$.

The obtained exponents $\nu$ and $\eta$ of the 3-state Potts
model seem to have lower accuracy than those of the Ising
model. The reason may be attributed to that we used the same
value of $m$ for both models. Since the Potts model has a
larger spin degree of freedom than the Ising model, a larger
size of matrix may be required to retain the  same order of
accuracy as the Ising case.

The authors would like to express their sincere thanks to
Y.~Akutsu for valuable discussions. T.~N. thank to
S.~R.~White for helpful discussions about the DMRG method. He
is also grateful to G.~Sierra for discussions about RSRG. The
present work is partially supported by a Grant-in-Aid from
Ministry of Education, Science and Culture of Japan. Most of
the numerical calculations were done by NEC SX-3/14R in
computer center of Osaka University.

\newpage

\section*{Figure Captions}

\begin{itemize}

\item[\bf Fig.~1.] Nearest neighbor spin correlation
function of the Ising model.

\item[\bf Fig.~2.] Spin polarization of the Ising model.

\item[\bf Fig.~3.] Two-parameter scaling function
$f(\xi(m)/N)$.

\item[\bf Fig.~4.] Nearest neighbor spin correlation
function of the 3-state Potts model.

\item[\bf Fig.~5.] Spin polarization of the 3-state Potts
model.

\end{itemize}
\end{document}